\documentclass[journal=jctcce,manuscript=article]{achemso}
\usepackage{achemso}
\setkeys{acs}{maxauthors=10, etalmode=truncate}
\usepackage{amssymb,amsfonts}
\usepackage{graphicx}
\usepackage{color}
\usepackage[table,dvipsnames]{xcolor}
\usepackage{amsmath}
\usepackage{bm}
\usepackage{braket}
\usepackage[version=4]{mhchem}
\usepackage{hyperref}
\usepackage{cleveref}
\usepackage{mciteplus}
\usepackage{setspace}
\usepackage{layout}
\usepackage{geometry}
\usepackage{siunitx}

\author{Kasper F. Schaltz}
\affiliation[dtu]{DTU Chemistry, Technical University of Denmark\\Kemitorvet Bldg. 206, 2800 Kgs. Lyngby, Denmark}
\author{Jonas Greiner}
\affiliation[dtu]{DTU Chemistry, Technical University of Denmark\\Kemitorvet Bldg. 206, 2800 Kgs. Lyngby, Denmark}
\author{Filippo Lipparini}
\affiliation[pisa]{Dipartimento di Chimica e Chimica Industriale, Universit\`a di Pisa\\Via G. Moruzzi 13, Pisa, 56124, Italy}
\author{Janus J. Eriksen}
\email{janus@dtu.dk}
\affiliation[dtu]{DTU Chemistry, Technical University of Denmark\\Kemitorvet Bldg. 206, 2800 Kgs. Lyngby, Denmark}

\title{Solvation Lies Within:\\Simulating Condensed-Phase Properties from Local Electronic Structures}

\begin{document}

\begin{abstract}

In transitions between different environmental settings, a molecular system inevitably undergoes a range of detectable changes, and the ability to accurately simulate such responses, e.g., in the form of shifts to molecular energies, remains an important challenge across physical chemistry. Based on an exact decomposition of total energies from Kohn--Sham density functional theory in a basis of spatially localized molecular orbitals, the present work outlines a robust protocol for sampling the effect of solvation within homogeneous condensed phases by focusing on perturbations to local electronic structures only. We report chemically intuitive results for binding energies of water, ethanol, and acetonitrile that all display fast convergence with respect to the bulk size. Results stay largely invariant with respect to the choice of basis set while reflecting differences in density functional approximations, and our protocol thus allows for a physically sound and efficient estimation of general effects related to bulk solvation. 

\end{abstract}

\newpage

\begin{figure}[ht]
\begin{center}
\includegraphics[width=\textwidth]{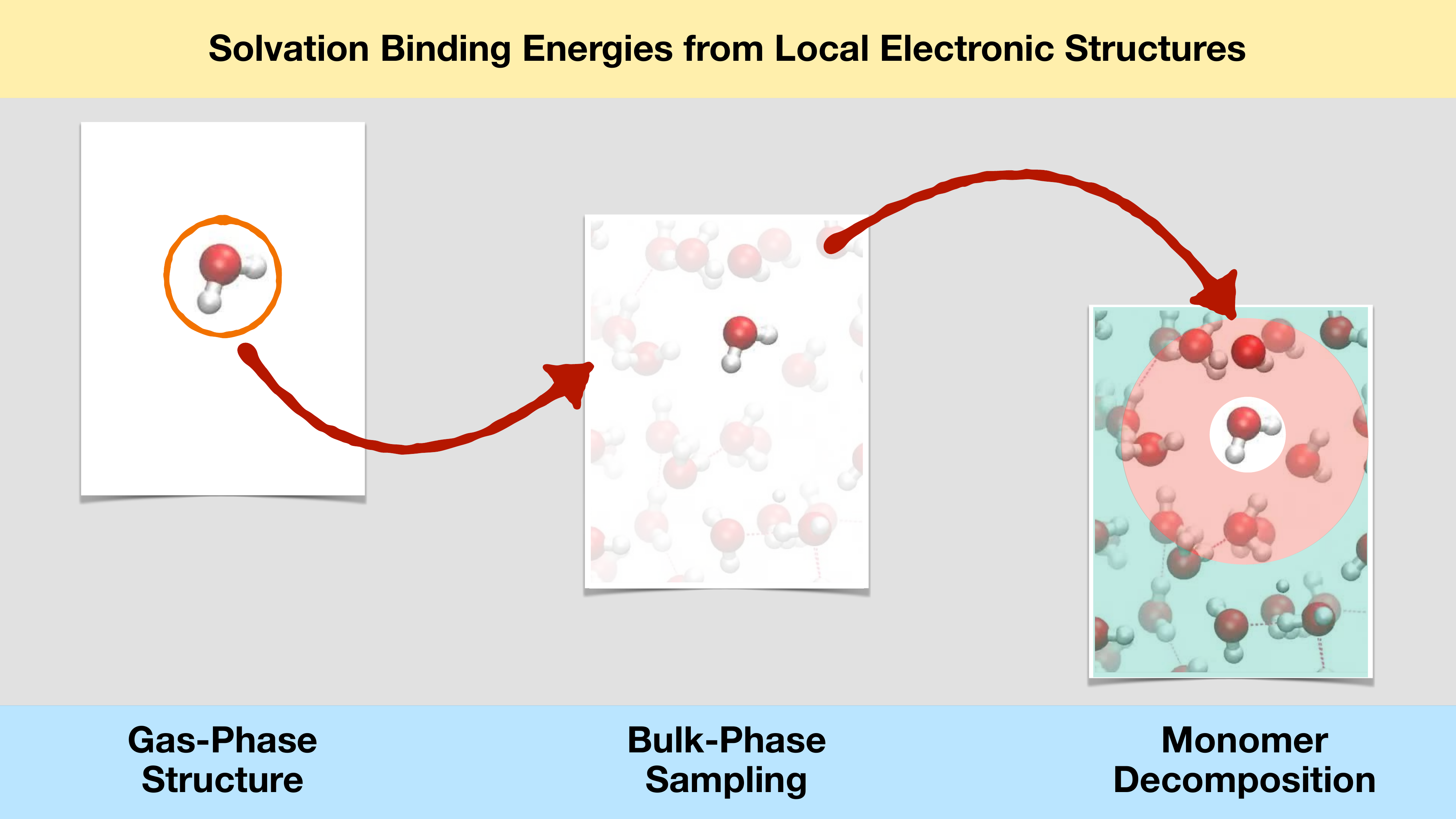}
\caption*{TOC graphic.}
\label{toc_fig}
\end{center}
\end{figure}

\newpage

\section{Introduction}\label{intro_sect}

The energy and charge distribution of a single molecule embedded within an environment are bound to differ from those of that same species in isolation, given how interactions with the surrounding medium will generally induce changes to these intrinsic properties. A move away from the gas phase will thus give rise to structural distortions but also perturbations to the local electronic structure of said molecule through both Coulombic interactions and possible polarization. Water, for instance, arguably the most common of liquids around, is also one of the most unusual on account of its ability to form hydrogen bonds \cite{franks2000water, Brini2017, Guillot2002, Scheiner1997}. Anomalies are extreme even under ambient and benign conditions, and a deep and unified understanding of water as a vapor, a liquid, or in any of its many crystalline and glassy forms at a microscopic level thus remains a key challenge within the field of physical chemistry~\cite{water_prop_chem_rev, Russo2014, Gallo2016}.\\

The present work proposes a novel protocol for unraveling and isolating individual contributions to such property shifts associated with a move from a vacuum to the bulk phase. The focus will be on perturbations to the local electronic structure of a central monomer only, as decomposed from a larger calculation on a bulk model. Our initial application will be concerned with water, given how its many peculiarities emerge from the structure and dynamics of its hydrogen bonding network, but we will also demonstrate the general nature of our approach by presenting results for liquid ethanol and acetonitrile. While outside the scope of the present work, the ability to probe the response to solvation and embedding of single molecules is interesting not just within homogeneous, condensed phases; rather, the developments outlined herein may also prove valuable in future studies of corresponding shifts to both extensive and intensive properties of chromophores in solutions or in various biological processes, e.g., for probing local phenomena in the vicinity of protein binding pockets~\cite{Marini2010,Ryde2016}.\\

The most rudimentary manner in which to simulate binding energies of homogeneous bulks would involve a calculation on a small cluster model, say, at the level of Kohn--Sham density functional theory~\cite{hohenberg1964inhomogeneous,kohn1965self,parr_yang_dft_book} (KS-DFT), and then averaging over the number of constituent monomers. The most obvious theoretical issue with such a methodology relates to artificial boundary effects, in that the outermost monomers are not subject to solvation to the same extent that the central monomers are. This error will only vanish slowly as the cluster is enlarged, thus incurring significant increases on the computational effort. Alternatively, one can simulate the effect of an environment onto a central moiety only by means of polarizable continuum models~\cite{Miertus1981, Cramer1999, continuum_review} (PCM), at the expense of being unable to account for strong local solute-solvent effects such as hydrogen bonds~\cite{mennucci_pcm_wires_2012, herbert_pcm_wires_2021}. As a remedy, one may instead represent the environment, that is, everything but a central monomer, by an analytical or derived force field~\cite{Ren2003,Nerenberg2018,Chipot2024}, by means of electrostatic and polarizable molecular mechanics (MM),~\cite{Warshel1976,Lin2007,QMMM_review_Senn,Olsen2010,Bondanza2020} or within more general quantum embedding frameworks where a manageable impurity region of interest, treated at a relatively high level of theory, gets embedded in a lower-level bath~\cite{Quantum_embedding_theories, Knizia2013, SeveroPereiraGomes2012}.\\

The formal exactness of all of these approaches can be naturally improved upon by including more than a single, central solute in the high-level region treated using quantum mechanics (QM), e.g., KS-DFT. However, this will necessarily reintroduce the earlier issue of how to probe changes to properties of just a single monomer of interest. To that end, it is worth mentioning that various subsystem-based approaches exist, all of which seek to reduce the computational scaling with system size by fragmentation into manageable parts~\cite{Gordon2012,Richard2012, Jacob2014,Herbert2019}. Such methods are, however, typically more concerned with the computation of total bulk energies or predicting the relative ranking of different cluster isomers than actual solvation shifts.\\

Instead, for the purpose of obtaining the energy of just a single moiety within an environment only, an altogether different strategy would be to extract this directly from a full QM calculation on a bulk model, in possible combination with an electrostatic and, potentially, polarizable MM region outside of this. To that end, atomic partitioning schemes based on the spatial locality of either native atomic (AOs) or molecular orbitals (MOs) have been proposed in the literature, each of which can potentially yield shifts in energies associated with the transition of an individual monomer from the gas and into the condensed phase by means of appropriate resummations. In the former of the two schemes, the so-called energy density analysis (EDA), every trace operation over products between integrals and the 1-electron reduced density matrix (1-RDM) of the KS-DFT energy functional is evaluated over only those AOs that are local to individual atoms~\cite{nakai_eda_partitioning_cpl_2002,nakai_eda_partitioning_cpl_2006,nakai_eda_partitioning_jcp_2007,nakai_eda_partitioning_ijqc_2009}. While orbital-invariant and independent of any population measures, results of such an AO-based decomposition will inevitably be sensitive to the composition of the underlying one-electron basis set ({\textit{vide infra}}). As an alternative, the full, spin-summed 1-RDM, $\bm{D}$, can be written as a sum over atom-specific counterparts, defined in a basis of spatially localized MOs, which are then traced with the integrals of the KS-DFT energy functional in lieu of $\bm{D}~\cite{eriksen_decodense_jcp_2020}$. Earlier works of ours have gathered substantial evidence in favor of an optimal combination of MOs and atomic weights to distribute these, namely, intrinsic bond orbitals (IBOs) and Mulliken populations derived in an intermediate, free-atom basis of intrinsic atomic orbital (IAOs)~\cite{ibo_iao,pm_orb_local_susi}. This scheme is optimal in the sense that it yields stable results for atomic energies upon a change of computational basis set, density functional approximation (DFA), and---not least---across identical functional groups embedded in different scaffolds~\cite{eriksen_elec_ex_decomp_jcp_2022,eriksen_local_prop_jctc_2023,eriksen_nn_qm7_decomp_jctc_2023,eriksen_trans_nn_mlst_2024,eriksen_pbc_decomp_jpca_2025}. We refer the reader to Sect. 1 of the supporting information (SI) for more details on the AO- and MO-based decomposition schemes used herein. \\

Before turning to actual bulk properties, it is informative to first study interactions between any two monomers under controlled conditions in order to understand how these might generalize from a model to a cluster. Using water as an example, we can thus start by systematically monitoring how the energies of the two monomers of a hydrogen-bonded water dimer change as the O--H$\cdots$O orientation angle is varied and the noncovalent bond gradually broken. When the two monomers orient themselves favorably, the local electronic structures associated with each of them become perturbed, causing the hydrogen bond to form. The question now is how the net lowering (stabilization) of the total energy of the dimer is reflected in the respective monomer energies? Traditionally, the monomer donating (accepting) a hydrogen to the bond is denoted as the {\textit{donor}} ({\textit{acceptor}}), but does this necessarily imply that the two monomers are correspondingly destabilized and stabilized, respectively, with respect to an optimized water molecule in the gas phase? As discussed above, PCM would not allow for the emergence of a spatially local hydrogen bond between the two, and determining the energies of both from individual calculations, in which the neighbor in both is represented by means of some flavor of quantum embedding, would neither be a sound approximation nor would there be any guarantee that the two monomer energies would add up to the total dimer energy. Of course, the latter condition is necessarily true if the dimer energy is simply averaged among the two monomers, but this result would give the obviously false impression that the two monomers are alike in the hydrogen-bonded complex.\\

\begin{figure}[ht]
    \centering
    \includegraphics[width=\textwidth]{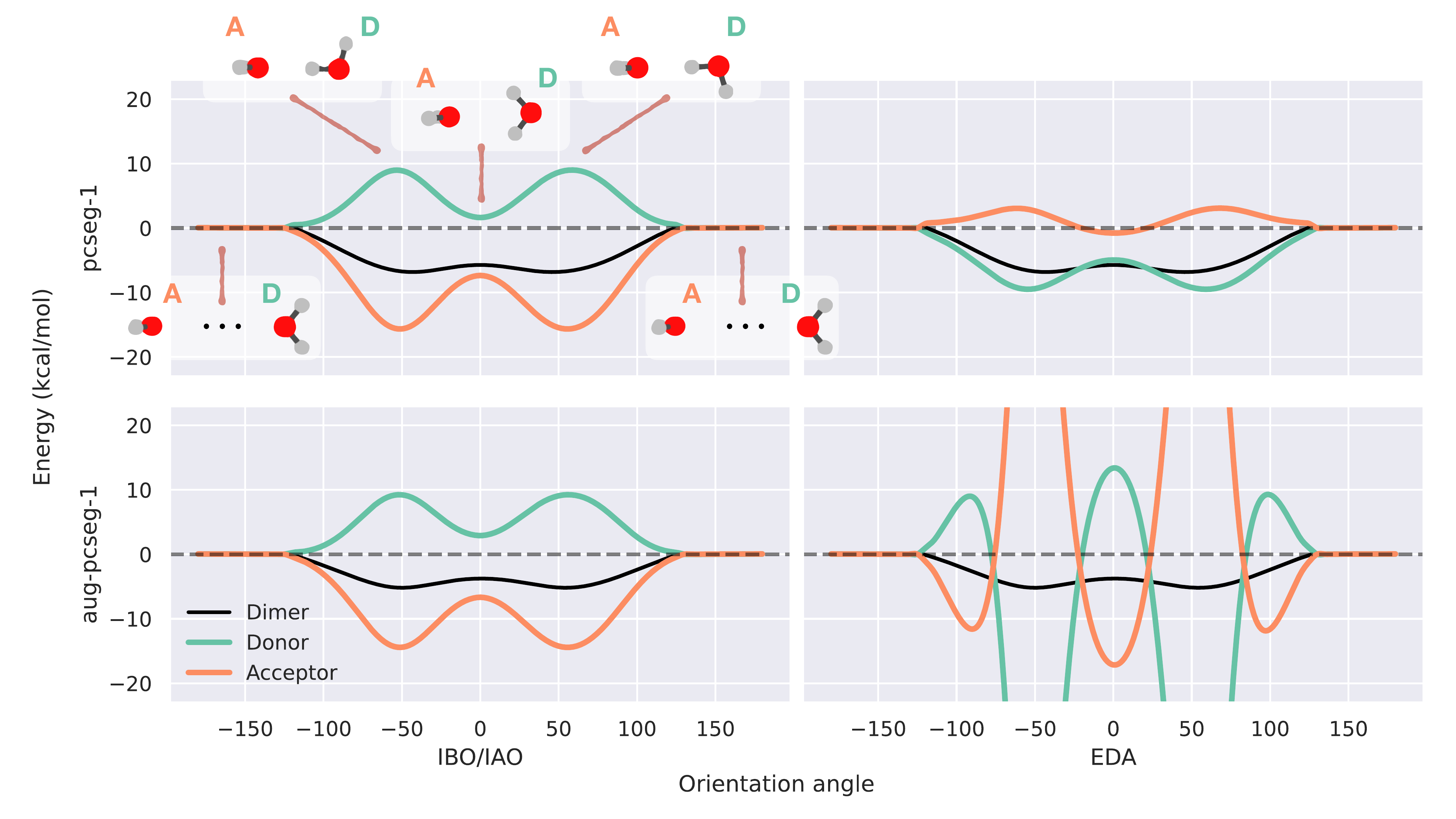}
    \caption{Changes to donor (D) and acceptor (A) energies in a water dimer upon forming and breaking hydrogen bonds. The monomer energies are decomposed using both MO- (left) and AO-based (right) schemes. Representative structures are provided in the upper-left panel.}
    \label{dimer_fig}
\end{figure}
As an alternative, the aforementioned AO- and MO-based decomposition schemes may be applied. These both partition the total dimer energy into atom-wise contributions, which may then readily be summed up into two monomer energies. The results of these schemes---using a combination of IBOs and IAO-based atomic weights for the latter---are presented in Fig. \ref{dimer_fig} for a constrained scan of the water dimer~\bibnote{The geometries for the dimer scan in Fig. \ref{dimer_fig} were obtained at the MP2/aug-cc-pVDZ~\cite{mp2_phys_rev_1934, dunning_4_aug} level of theory in \texttt{CFOUR}~\cite{cfour_paper}. All structural parameters were optimized except for the relative orientation of one water molecule with respect to the other.}. The decompositions were performed at the $\omega$B97M-V/(aug-)pcseg-1~\cite{wb97mv,pc_basis} level of theory using the implementations in {\texttt{decodense}}~\cite{decodense}.\\

We begin by discussing the MO-based (IBO/IAO) results on the left-hand side of Fig. \ref{dimer_fig}. In forming either of the two identical hydrogen bonds, the monomer donating its hydrogen to the bond experiences an increase in energy while the acceptor unit gets stabilized to an even greater degree, thus resulting in an overall stabilization of the dimer. In the intermediate regime around an orientation angle of $0 \si{\degree}$, the dimer is still stabilized on account of two partial hydrogen bonds; here, the acceptor too remains stabilized with respect to the gas phase, whereas the energy of the donor is only marginally changed. These observations are found to hold true in any of the two basis sets, that is, even upon adding diffuse functions.\\

In the AO-based EDA results, on the other hand, the picture in the pcseg-1 basis set is observed to be the exact opposite of that discussed above. Namely, the acceptor is destabilized (except for whenever the two bonds are only partial), while the donor is stabilized at all angles. However, in the augmented basis, results are both exaggerated and bear no resemblance to those in the smaller basis. This discrepancy between the results is a direct consequence of the strong sensitivity of EDA on the composition of the basis set at hand alluded to earlier, a variation which ultimately renders such AO-based decompositions less suitable for the present purpose of probing local electronic structures than the corresponding IBO/IAO scheme. For this reason, we will only concern ourselves with the latter herein.\\

Having looked at how to infer monomer energies from a model system, the remainder of this work will be concerned with studying how to simulate solvation processes from such decompositions of KS-DFT. To what extent does the energy associated with an individual monomer change whenever this becomes suspended in a condensed phase, how to account for temperature effects and conformational sampling across different bulk models, and how far do these bulks need to extend to compute converged energy shifts? In the process, we will compare decomposed binding energies with alternative quantities derived from total bulk energies as well as available enthalpies of vaporization at room temperature, the experimental property which is arguably closest in nature to what we are simulating in the present study.\\

The outline of our work is as follows. In Sect. \ref{theory_sect}, we present a new protocol for determining solvation binding energies within homogeneous condensed phases based on the MO-based decomposition (IBO/IAO) of Fig. \ref{dimer_fig}. Sect. \ref{comp_detail_sect} next provides computational details behind the results for the water, ethanol, and acetonitrile systems to follow in Sect. \ref{res_sect}, before Sect. \ref{sum_disc_sect} presents a summary alongside some conclusive discussions and a look at future applications.

\section{Theory}\label{theory_sect}

Our protocol for simulating solvation properties is outlined in Sect. \ref{solvation_subsect} whereas aspects concerned with sampling and uncertainty measurements are discussed in Sect. \ref{sampling_subsect}. Alternative expressions for binding energies involving the same calculations are discussed in Sect. \ref{alternatives_subsect}.

\subsection{Binding Energies}\label{solvation_subsect}

\begin{figure}[ht]
    \centering
    \includegraphics[width=\textwidth]{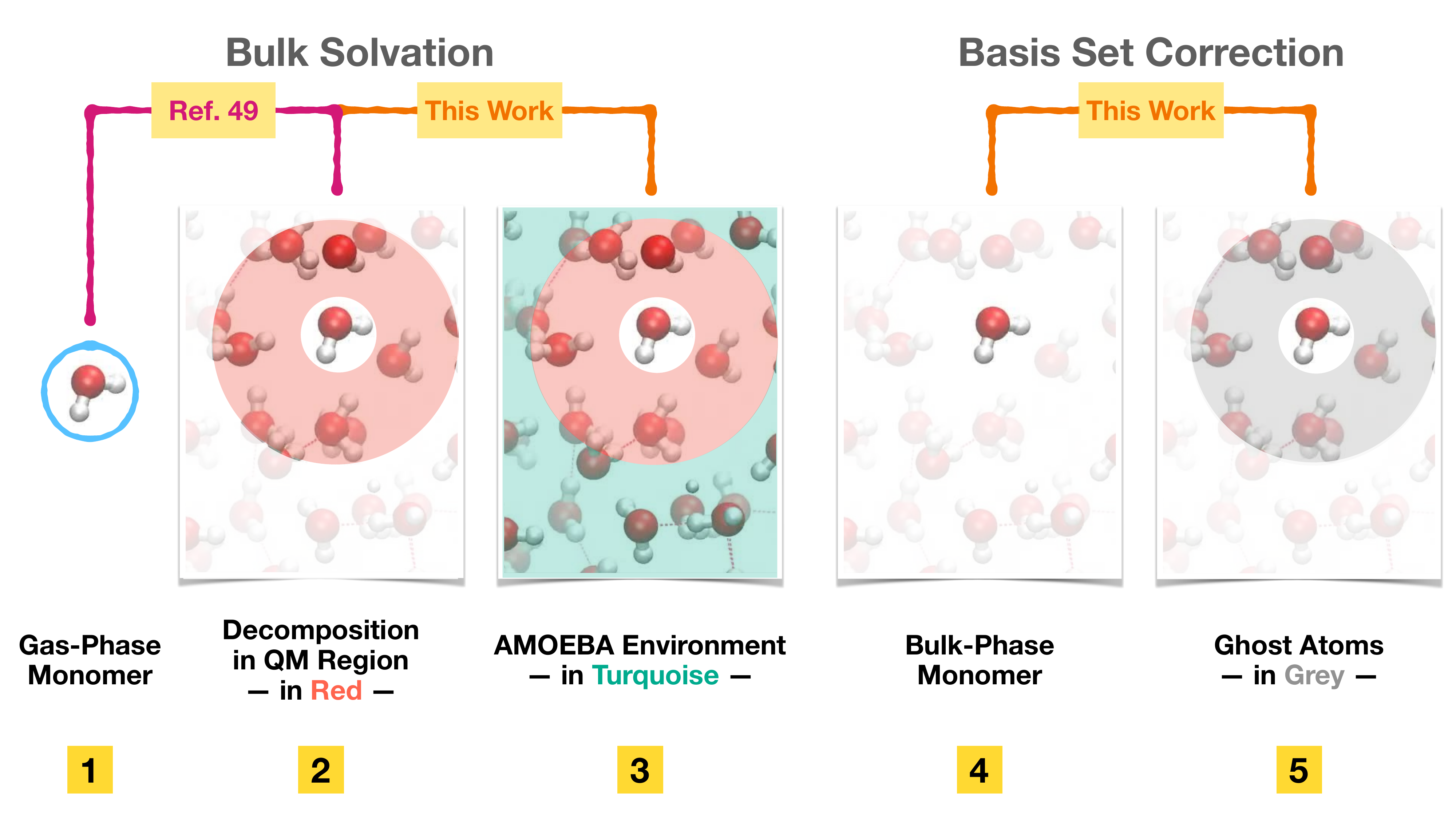}
    \caption{Individual steps and calculations involved in the solvation protocol of the present work. For comparison, the protocol of Ref. \citenum{eriksen_local_condensed_phase_jpcl_2021} has also been highlighted in dark magenta.}
    \label{protocol_fig}
\end{figure}
The transition from a model dimer to a bulk phase warrants a number of additional considerations concerned with discretization, sampling, structural relaxation, alongside the possible treatment of environment effects when one wants to simulate an overall solvation process. In an earlier study, primarily focused on solvation shifts to the dipole moment of water, results were computed on par with Fig. \ref{dimer_fig}, that is, as the difference in energy between a solvated monomer suspended in a bulk phase and an isolated monomer at equilibrium geometry in the gas phase~\cite{eriksen_local_condensed_phase_jpcl_2021}. In the context of the illustration in Fig. \ref{protocol_fig}, this corresponds to the difference in energy between panel $\bm{2}$, subject to sampling in the condensed phase, and panel $\bm{1}$, albeit at a temperature of $T=0$ K. We will now outline how to improve upon this na{\"i}ve protocol.\\

The first thing to notice is that any solvation process in a homogeneous phase will strictly occur at a finite temperature, typically under ambient conditions. To allow for computed quantities to be compared to experiment, e.g., measured enthalpies of vaporization, one needs thus to account for not only bulk effects but also vibrational and rotational contributions. In here, our reference energy, $E^{\text{vac}}$, is computed at $T = 298$ K by adding a thermal correction to the equilibrium energy computed from Hessian information (the standard ideal gas/rigid rotor/harmonic oscillator approximation~\cite{Swendsen2019}, but omitting the zero-point vibrational energy correction~\bibnote{In contrast to the temperature-dependent vibrational contributions, the zero-point energy correction is not recovered through the MD sampling in our protocol and has therefore been omitted.}). Next, the energy of a central monomer of interest ($\mathcal{K}$) surrounded by $n$ neighbours in a bulk, $\mathcal{E}^{(n)}_{\mathcal{K}}$, decomposed from the corresponding total energy, $E^{(n)}$, and computed using the calculations depicted in either of panels $\bm{2}$ or $\bm{3}$ of Fig. \ref{protocol_fig} ({\textit{vide infra}}), will necessarily be spanned in a significantly larger set of basis functions than that which is available to the monomer in the gas phase, $E^{\text{vac}}$. To account for this difference in discretization, but also the structural relaxation taking place in moving from a vacuum to the bulk, we will here add to our results a correction involving two separate calculations, namely, one of the same monomer ($\mathcal{K}$) as used in panels $\bm{2}$ or $\bm{3}$, suspended in the bulk but in the absence of any neighbors, $E^{(0)}_{\mathcal{K}}$, and the same monomer in the presence of $n$ surrounding monomers replaced by ghost atoms and their basis functions (G), $E^{(0,\text{G})}_{\mathcal{K}}$. This correction thus lends itself to a counterpoise strategy for ameliorating basis set superposition errors (BSSEs)~\cite{BSSE_counterpoise, Duijneveldt1994}.\\

Our final expression for computing what we denote as {\textit{binding energies}} thus reads as follows:
\begin{align}
\Delta E^{(n)}_{\mathcal{K}} = \mathcal{E}^{(n)}_{\mathcal{K}} - E^{\text{vac}} + (E^{(0)}_{\mathcal{K}} - E^{(0,\text{G})}_{\mathcal{K}}) \ . \label{solv_energy_eq}
\end{align}
The snapshots of panels $\bm{2}$--$\bm{5}$ in Fig. \ref{protocol_fig} are drawn from the same distributions, cf. Sect. \ref{sampling_subsect}.\\

With respect to Ref. \citenum{eriksen_local_condensed_phase_jpcl_2021}, the possible treatment of environment effects outside the QM region shaded in red in Fig. \ref{protocol_fig} has been carefully reconsidered (cf. panel $\bm{3}$). Instead of employing a simple point-charge model of the MM region, we here make use of the AMOEBA force field, consisting of both electrostatic, polarizable, and dispersion-like terms, which all affect the QM region~\cite{amoeba}. In particular, both electrostatic and polarization terms enter the QM molecular Hamiltonian, and the QM and polarization equations are solved self-consistently to account for mutual polarization effects~\cite{loco2016,loco2017,loco2019b,Lipparini2019,Nottoli2023}. Beyond a much more sophisticated environment potential, the actual use of this in the context of Fig. \ref{protocol_fig} has also been modified. Rather than using the contribution from the central monomer to the entire QM/AMOEBA energy as $\mathcal{E}^{(n)}_{\mathcal{K}}$ in Eq. \ref{solv_energy_eq}, which will inevitably suffer from a slow convergence with the extent of the main QM region due to electrostatics, we here account only for the polarization of the environment region onto the density matrix of a central monomer. Namely, the effect of the AMOEBA force field is included by replacing the MOs of panel $\bm{2}$ in Fig. \ref{protocol_fig} with those obtained in a corresponding QM/MM calculation, but retaining only QM terms to the energy, $\mathcal{E}^{(n)}_{\mathcal{K}}$.\\

Finally, on par with electrostatics, AMOEBA includes a treatment of dispersion, which, in any homogeneous liquid, is both important and bound to converge rather slowly with respect to outwards distance from a central monomer. However, when using a DFA that too treats long-range interactions, such as $\omega$B97M-V~\cite{wb97mv}, which accounts for non-local correlation by means of the VV10 functional~\cite{vv10}, it arguably makes sense to add part of the dispersion contributions from AMOEBA in order to balance these in the overall treatment of environment effects. In AMOEBA, van der Waals (vdW) interactions are treated by a buffered 14--7 Lennard-Jones correction (cf. SI). For our purposes, interest is only on how the pairwise interactions between a central monomer and all MM atoms affect the former, and we thus scale this contribution by an overall factor of $0.5$. While by no means consistent, such an {\textit{ad hoc}} correction still yields a more balanced treatment of long-range interactions than if not included. However, whenever the DFA of choice is not augmented by dispersion corrections (say, when using the B3LYP~\cite{becke_b3lyp_functional_jcp_1993,frisch_b3lyp_functional_jpc_1994} or MN15~\cite{Yu2016_MN15} functionals, both of which will be assessed in the present study), this scaled contribution is not included in the calculation of $\mathcal{E}^{(n)}_{\mathcal{K}}$.

\subsection{Sampling}\label{sampling_subsect}

As is obvious from the results in Fig. \ref{dimer_fig}, the (de)stabilization of individual monomers in a cluster is bound to vary substantially for different bulk configurations. For sampling the dynamics within the condensed phase, it is fundamental to sample a statistically significant distribution, e.g., to account for local effects, such as hydrogen bonding, when modeling the partition function of the total system. MD approaches essentially involve a thermodynamic integration, in that time-wise equidistant
samplings are used in an attempt at representing the exact, temperature-dependent system dynamics, and the prevalence of a specific configuration is represented by equally-weighted statistics. On the other hand, one can opt for a more static approach where one randomly generates and optimizes snapshots, subject to a nonuniform weighting of the different configurations. In the present study, we have decided upon the former strategy but we comment further on the potential use of the latter in Sect. \ref{sum_disc_sect}.\\

At the same time, given how we are decomposing the energy of a central water monomer from a larger QM or QM/MM calculation for each of these snapshots, the size of the QM region will also need to be properly converged, either in terms of outwards distance or, as we will do in the course of the present work, in terms of the number of QM monomer neighbors.\\

\begin{figure}[ht]
    \centering
    \includegraphics[width=\textwidth]{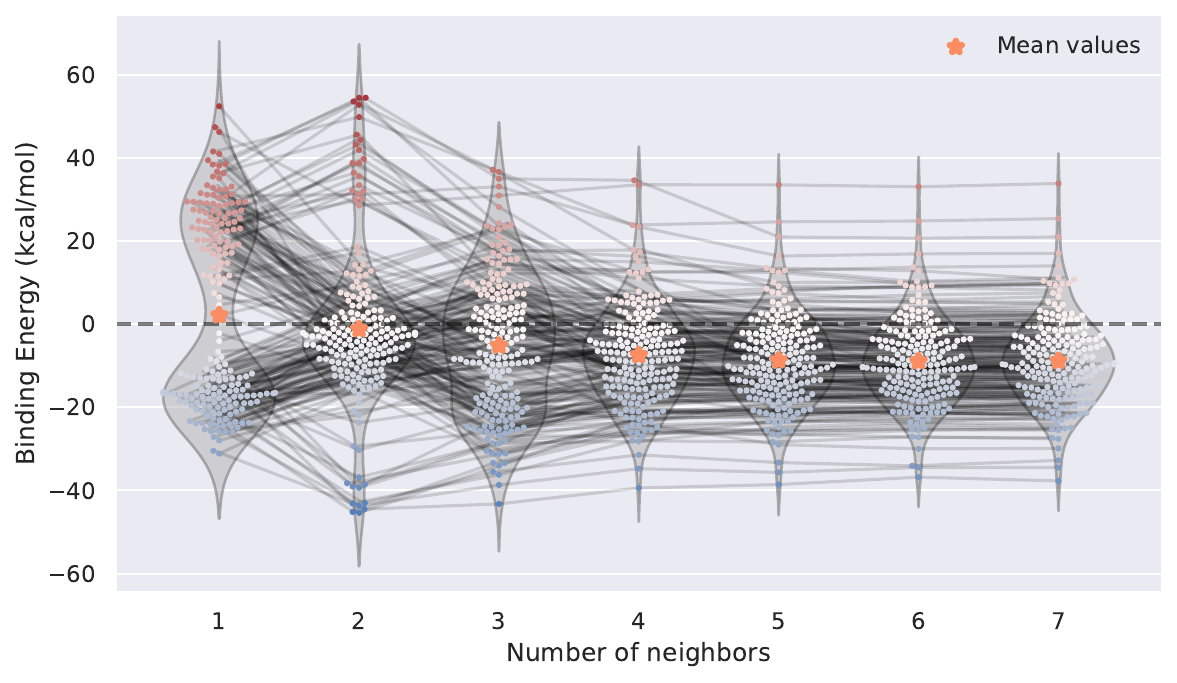}
    \caption{Convergence of binding energies ($\Delta E^{(n)}_{\mathcal{K}}$) obtained using Eq. \ref{solv_energy_eq} for 200 individual snapshots of bulk water. The level of theory is $\omega$B97M-V/aug-pcseg-1 and the QM/AMOEBA-based protocol depicted in panel $\bm{3}$ of Fig. \ref{protocol_fig} is used in all calculations.}
    \label{swarmplot_mbpol_fig}
\end{figure}
The evolution of the binding energy associated with individual snapshots of water drawn from a 256-monomer MD sampling is shown in Fig. \ref{swarmplot_mbpol_fig}~\bibnote{Neighboring monomers are ranked on the basis of the shortest distance between any one atom and any of those constituting the central monomer of interest.}. The results clearly show how, upon inclusion of just one neighbor in the QM region, the two separate distributions of Fig. \ref{dimer_fig} are observed with a significant gap in-between, while a unimodal distribution soon emerges as the QM region is further enlarged. In general, only a small relative number of snapshots represent a central water monomer acting as donor or acceptor in two successive strong hydrogen bonds, leading to additional (de)stabilization in moving from $n=1$ to $2$, and beyond at most $5$ waters surrounding the central monomer, no significant statistical change is observed. In Sect. \ref{res_sect}, we will numerically confirm this convergence with respect to cluster size.\\

The statistical uncertainty on mean binding energies from a finite number of snapshots in our analyses, $\mathcal{N}$, can be determined by the technique of bootstrapping~\cite{bootstrap}. Here, distributions are resampled with replacement and means calculated over these, before the uncertainty is computed as the 95th percentile of the absolute difference between the new means and that of the original distribution. The standard error of the sample mean decreases proportionally to the inverse square root of $\mathcal{N}$. Consequently, reducing the sampling error by half requires a fourfold increase in sample size, highlighting diminishing returns and underscoring the importance of selecting a sample size that balances statistical precision with practical constraints. As demonstrated in Fig. S1 of the SI, we find $\mathcal{N}=200$ to be a pragmatic total number of snapshots, i.e., as a compromise between computational effort and accuracy.

\subsection{Alternative Protocols}\label{alternatives_subsect}

\begin{figure}[ht]
    \centering
    \includegraphics[width=\textwidth]{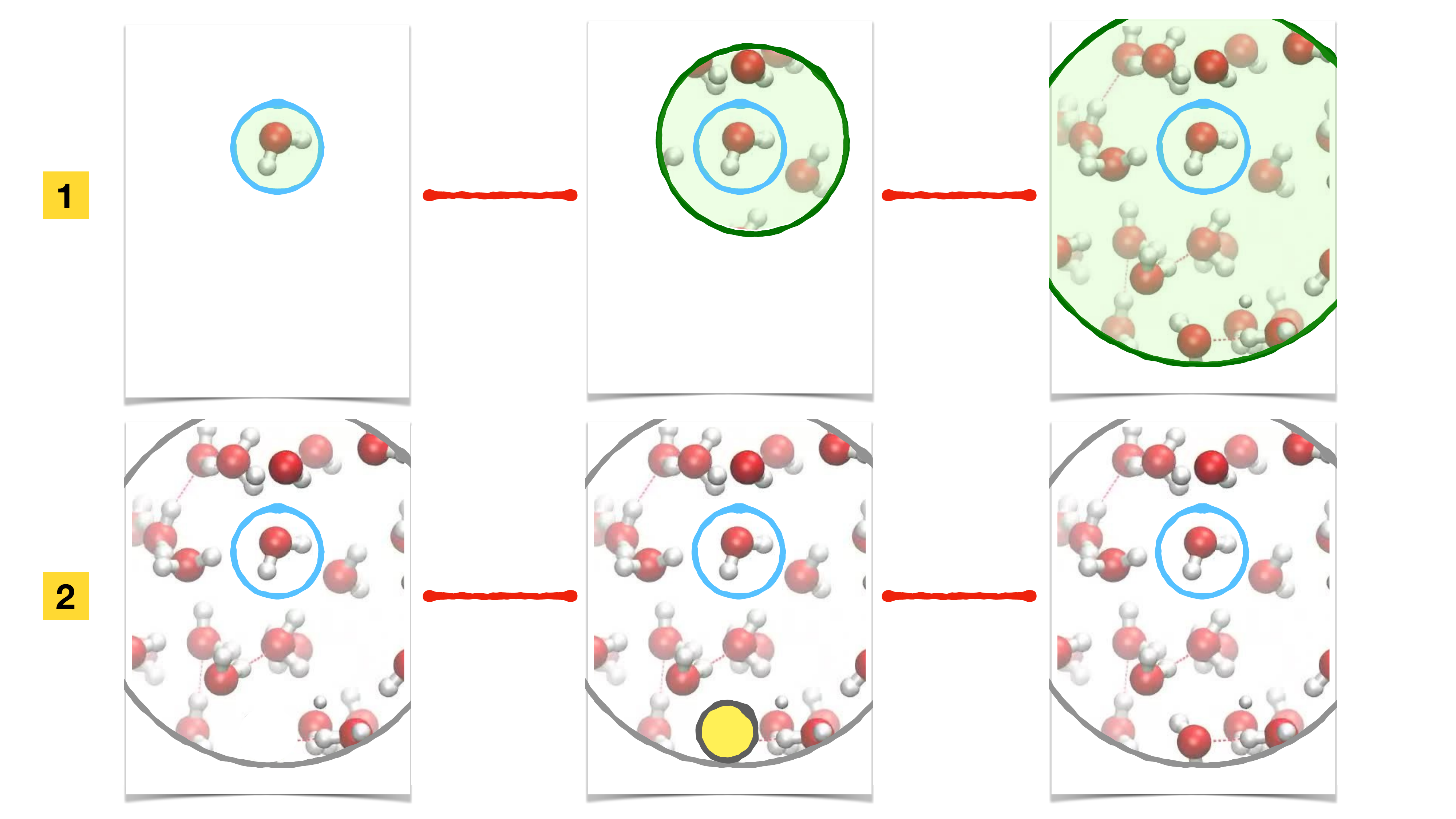}
    \caption{Two alternatives to Eq. \ref{solv_energy_eq}. Average ($\bm{1}$) and incremental ($\bm{2}$) solvation energy shifts are derived from successively larger bulk models according to Eqs. \ref{average_energy_eq} and \ref{inc_energy_eq}, respectively.}
    \label{comparison_eqs_fig}
\end{figure}
On the basis of the individual calculations behind Eq. \ref{solv_energy_eq}, there exist at least two alternative ways in which to derive similar binding energies. As touched upon in Sect. \ref{intro_sect}, the arguably most elementary protocol infers monomer energies by simply averaging those for the bulk:
\begin{align}
\Delta E^{(n)}_{\text{av}} = \frac{E^{(n)}}{n+1} - E^{\text{vac}} \ . \label{average_energy_eq}
\end{align}
Due to obvious boundary effects, the energy shifts computed from Eq. \ref{average_energy_eq} are expected to converge rather slowly upon an increase of the extent of the QM region ($n+1$). Alternatively, one can estimate the effect of solvation by monitoring the incremental shifts instead:
\begin{align}
\Delta E^{(n)}_{\text{inc}} = (E^{(n)} - E^{(n-1)}) - E^{\text{vac}} \ . \label{inc_energy_eq}
\end{align}
Eqs. \ref{average_energy_eq} and \ref{inc_energy_eq} are schematically compared in panels $\bm{1}$ and $\bm{2}$ of Fig. \ref{comparison_eqs_fig}, respectively (cf. also the present protocol in Fig. \ref{protocol_fig}). Among the two, the latter formalism is the only that bears some (faint) resemblance to how classical solvation thermodynamics is typically partitioned into two components: first, an initial, nontransient cavity is formed in the full bulk, followed by the introduction of an interacting solute into this. When formulated in terms of free rather than the internal energies of the present work, the first step thus describes the reversible work needed to open up an appropriate bulk cavity, while the second expresses the resulting binding interactions~\cite{ben_naim_book}. In panel $\bm{2}$ of Fig. \ref{comparison_eqs_fig}, no cavity is ever explicitly formed, and the protocol proceeds simply by successively adding nearest neighbours one at a time, cf. Eq. \ref{inc_energy_eq}.

\section{Computational Details \label{comp_detail_sect}}

For the initial case of water in Sect. \ref{water_res_subsect}, we assess three methods of sampling bulk dynamics, all differing in the compromise between expected accuracy and computational efficiency, with an eye on the study of the ethanol and acetonitrile systems to follow in Sect. \ref{ethanol_acetonitrile_res_subsect}. Radial distribution functions of the noncovalent bond distances are presented in Fig. S2 of the SI.\\

The MB-Pol model developed by Paesani {\textit{et al.}} has emerged as today's state-of-the-art for the sampling of liquid and solid-state water~\cite{Babin2013,paesani_mbpol_chem_rev_2016}. In the present study, snapshots were generated using the MB-Pol(2023)/2c3b4a version in the {\texttt{MBX}} plugin to {\texttt{LAMMPS}}~\cite{mbpol2023,mbx,LAMMPS}. The bulk system consisted of a total of 256 water molecules within a cubic, periodic box, simulated in an $NpT$ ensemble at $T=298$ K and $p=1$ bar. The simulation timestep was set to 0.5 fs, and, following a suitable equilibration period, snapshots were sampled at intervals of 1.5 ps.\\

As a second option, we also performed {\textit{ab initio}} molecular dynamics (AIMD) in a basis of Gaussian plane waves using {\texttt{CP2K}}~\cite{VandeVondele2003,Hutter2013,Kuehne2020}. The revPBE functional was employed in combination with Grimme's D3 dispersion corrections~\cite{Perdew1996, Zhang1998, Grimme2010, Grimme2011}, and core electrons were treated using standard GTH pseudopotentials~\cite{Goedecker1996, Hartwigsen1998, Krack2005}, while valence electrons were expanded in a DZVP-MOLOPT-SR-GTH basis set with a plane wave cutoff of $400$ Ry~\cite{VandeVondele2007}. Periodic boundary conditions were applied to a simulation cell that reproduced the density of water at room temperature. The system was propagated in an $NVT$ ensemble at $T=300$ K using a generalized Langevin equation thermostat with a timestep of $1.0$ fs~\cite{Ceriotti2009, Ceriotti2009a}, sampling individual snapshots every $0.5$ ps.\\

Finally, as a significantly more cost-efficient alternative, we also employed the new equivariant smooth energy network small conserving model, {\texttt{eSEN-sm-cons}}, a machine-learned interatomic potential trained on the recent Open Molecules 2025 dataset~\cite{Fu2025,Levine2025}. For all three systems, snapshots were generated every $1.0$ ps using a timestep of 1.0 fs (at $T=298$ K).\\

All QM and QM/AMOEBA calculations of the present work have been performed using the {\texttt{PySCF}} package~\cite{pyscf2018,pyscf2020,libxc}, its interface with the {\texttt{OpenMMPol}} library~\cite{ommp}, and the {\texttt{decodense}} code~\cite{decodense}.

\section{Results}\label{res_sect}

We start by assessing the protocol presented in Sect. \ref{solvation_subsect} for the case of liquid water in Sect. \ref{water_res_subsect}. Results will be compared to those obtained using the alternative protocols of Sect. \ref{alternatives_subsect}, and we will furthermore also compare the different methods of sampling the bulk dynamics, cf. Sect. \ref{comp_detail_sect}. Finally, results for ethanol and acetonitrile will be discussed in Sect. \ref{ethanol_acetonitrile_res_subsect}.

\subsection{Water}\label{water_res_subsect}

\begin{figure}[ht!]
    \centering
    \includegraphics[width=\textwidth]{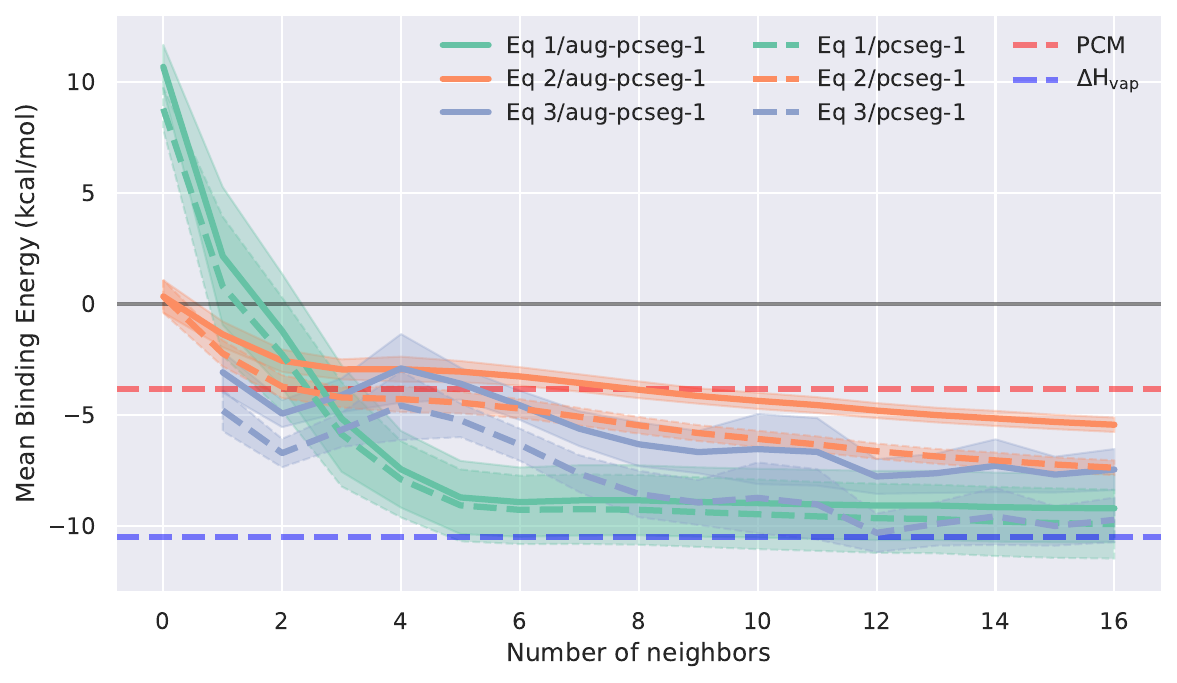}
    \caption{Comparison of the three different protocols for computing the binding energy of water studied herein, namely, the present MO-based, decomposed approach in Eq. \ref{solv_energy_eq}, $\Delta E^{(n)}_{\mathcal{K}}$, the averaged approach in Eq. \ref{average_energy_eq}, $\Delta E^{(n)}_{\text{av}}$, and the incremental approach in Eq. \ref{inc_energy_eq}, $\Delta E^{(n)}_{\text{inc}}$. All the results using Eqs. \ref{solv_energy_eq}--\ref{inc_energy_eq} are obtained at the $\omega$B97M-V/(aug-)pcseg-1 level of theory, and those based on Eq. \ref{solv_energy_eq} are obtained from QM/AMOEBA calculations (cf. panel $\bm{3}$ of Fig. \ref{protocol_fig}).}\label{comparison_eqs_fig_water}    
\end{figure}
We start by comparing results for mean binding energies of water obtained using the present protocol in Eq. \ref{solv_energy_eq} to results derived instead using the average and incremental protocols in Eqs. \ref{average_energy_eq} and \ref{inc_energy_eq}, respectively. Fig. \ref{comparison_eqs_fig_water} compares results at the $\omega$B97M-V/(aug-)pcseg-1 level of theory; those derived from Eq. \ref{solv_energy_eq} are obtained in the presence of an AMOEBA environment outside the QM region (cf. panel $\bm{3}$ of Fig. \ref{protocol_fig}), while the corresponding results based on Eqs. \ref{average_energy_eq} and \ref{inc_energy_eq} are not. For reference, earlier studies of water based on Eq. \ref{average_energy_eq}---formulated around either KS-DFT or fragment-based perturbation theory---have reported average binding (or stabilization) energies with respect to optimized gas-phase structures of around $-9$ kcal/mol, requiring a combination of relatively large QM clusters and the use of extended basis sets~\cite{Maheshwary2001,Pruitt2012}.\\

A number of key observations can be made on the basis of the results in Fig. \ref{comparison_eqs_fig_water}. First, the three different expressions for the binding energy of water fail to yield results in satisfactory agreement with one another even when using QM regions comprised of a total of $16$+$1$ water monomers. Second, in terms of numerical convergence, the profile of the results derived from Eq. \ref{solv_energy_eq} is observed to be significantly smoother and more stable than those obtained using either of Eqs. \ref{average_energy_eq} and \ref{inc_energy_eq}; only marginal changes to the mean binding energy are observed at $n \geq 5$. Third, the variance in the results is observed to be the smallest for the results obtained using Eq. \ref{average_energy_eq}, which is testament to the fact that local variations in-between different snapshots are practically indistinguishable in this protocol where monomer energies are obtained simply by averaging over bulk energies. The bootstrapping uncertainties on the mean energies derived from Eqs. \ref{solv_energy_eq} and \ref{inc_energy_eq} (Sect. \ref{sampling_subsect}) are both substantially larger, reflecting the differences in local electronic structure of a central water monomer discussed earlier (cf. Fig. \ref{dimer_fig}) and how these depend crucially on its network of surrounding hydrogen bonds.\\

The invariance under a change of basis set is also observed to be the smallest in the binding energies based on Eq. \ref{solv_energy_eq}. This point is further emphasized in the results of Fig. S3 of the SI, in which $\omega$B97M-V results in (aug-)pcseg-$x$ ($x=0,1$) basis sets are compared. Here, the pcseg-0 basis set, in particular, is observed to be too limited in composition and size to be applicable for the present purpose, with binding energies failing to converge satisfactorily even for more extended QM regions. The reason for this is the large magnitude and slow convergence of the BSSE correction in this basis set, cf. Fig. S5 of the SI. However, even standard augmentation by diffuse functions (aug-pcseg-0) is observed to remedy this shortcoming to a large extent, although better still is to move to the pcseg-1 basis set and, in particular, the aug-pcseg-1 variant. Given how all other contributions to Eq. \ref{solv_energy_eq} are found to converge similarly in the four different basis sets, it is thus the improved balance in the discretization of the central monomer of interest in moving from vacuum to the bulk which is identified as being key to the rapid convergence of the mean binding energy in Fig. \ref{comparison_eqs_fig_water}.\\

In the case of water, the variance with the choice of exchange-correlation ($xc$) functional is also not observed to be too severe. In Fig. S4, results in a pcseg-1 basis set are compared using B3LYP, MN15, and $\omega$B97M-V, as three representative examples of different, yet popular DFAs. Overall, the three qualitatively agree with one another, both in terms of convergence profiles and their asymptotic limits, with the latter results (based on $\omega$B97M-V) observed to agree the closest with the experimental enthalpy of vaporization at room temperature, denoted in Fig. \ref{comparison_eqs_fig_water} by a dashed horizontal line in blue ($\Delta \text{H}_{\text{vap}}$). The improvement over a simple PCM estimate at the $\omega$B97M-V/aug-pcseg-1 level of theory is similarly obvious.\\

\begin{figure}[ht!]
    \centering
    \includegraphics[width=\textwidth]{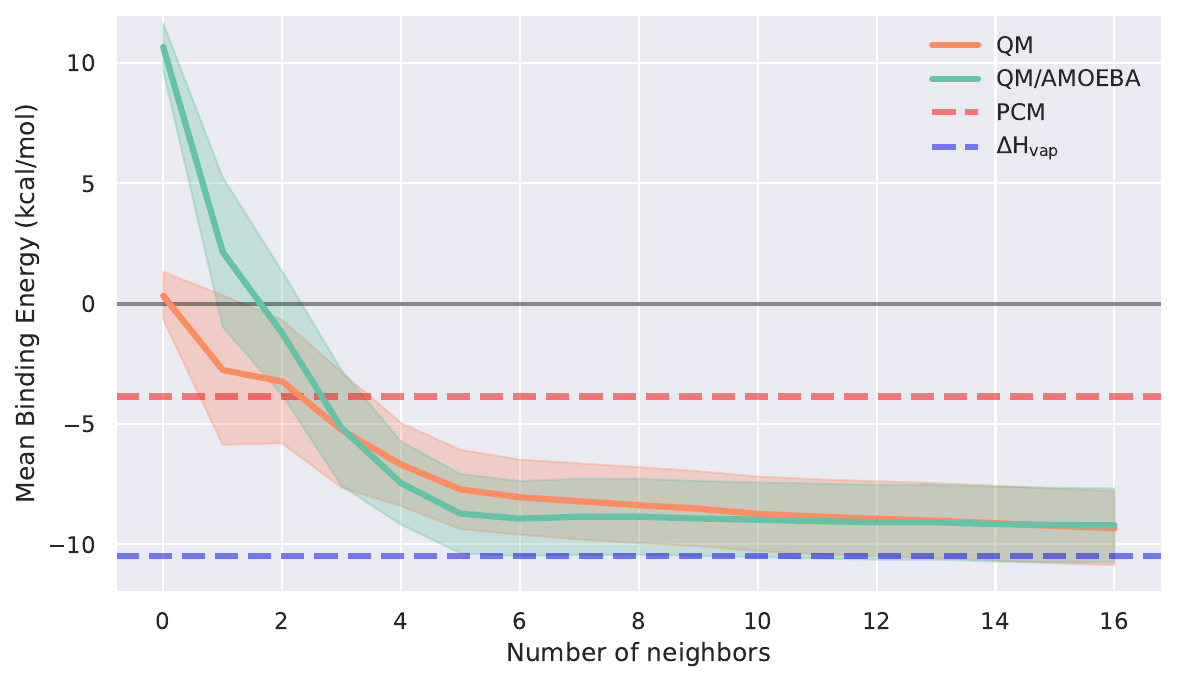}
    \caption{Comparison of QM- and QM/AMOEBA-based approaches to computing the decomposed energy, $\mathcal{E}^{(n)}_{\mathcal{K}}$, of Eq. \ref{solv_energy_eq} for water. The two approaches correspond to panels $\bm{2}$ and $\bm{3}$ of Fig. \ref{protocol_fig}, respectively. For both sets of results, the level of theory is $\omega$B97M-V/aug-pcseg-1.}\label{comparison_panels_fig}  
\end{figure}
Next, in Fig. \ref{comparison_panels_fig}, we assess the added benefits of computing $\mathcal{E}^{(n)}_{\mathcal{K}}$ of Eq. \ref{solv_energy_eq} by means of the QM/AMOEBA-based calculation in panel $\bm{3}$ of Fig. \ref{protocol_fig} rather than the bare QM calculation in panel $\bm{2}$. Although the asymptotic limits are the same, results are indeed observed to be accelerated by adding a structured environment outside the main QM region. Results are observed not to change for $n \geq 5$, even though the result obtained with just the central water monomer embedded within AMOEBA is more than $10$ kcal/mol off the corresponding result obtained without an environment description. In the case of $\omega$B97M-V, we recall how the effect of adding AMOEBA to our protocol relates not just to using MOs obtained in the presence of a polarizable environment but also to adding a scaled vdW contribution between QM and MM atoms to the VV10 treatment of dispersion inside the QM region. The explicit effect of adding these (basis set-independent) vdW contributions are shown in Fig. S5 of the SI. Collectively, we thus conclude that the convergence observed at the $\omega$B97M-V/aug-pcseg-1 level of theory obviously manifests for the right reasons, given how convergence is observed not only for the total sum of the contributions, but also for the individual ones.\\

\begin{figure}[ht!]
    \centering
    \includegraphics[width=\textwidth]{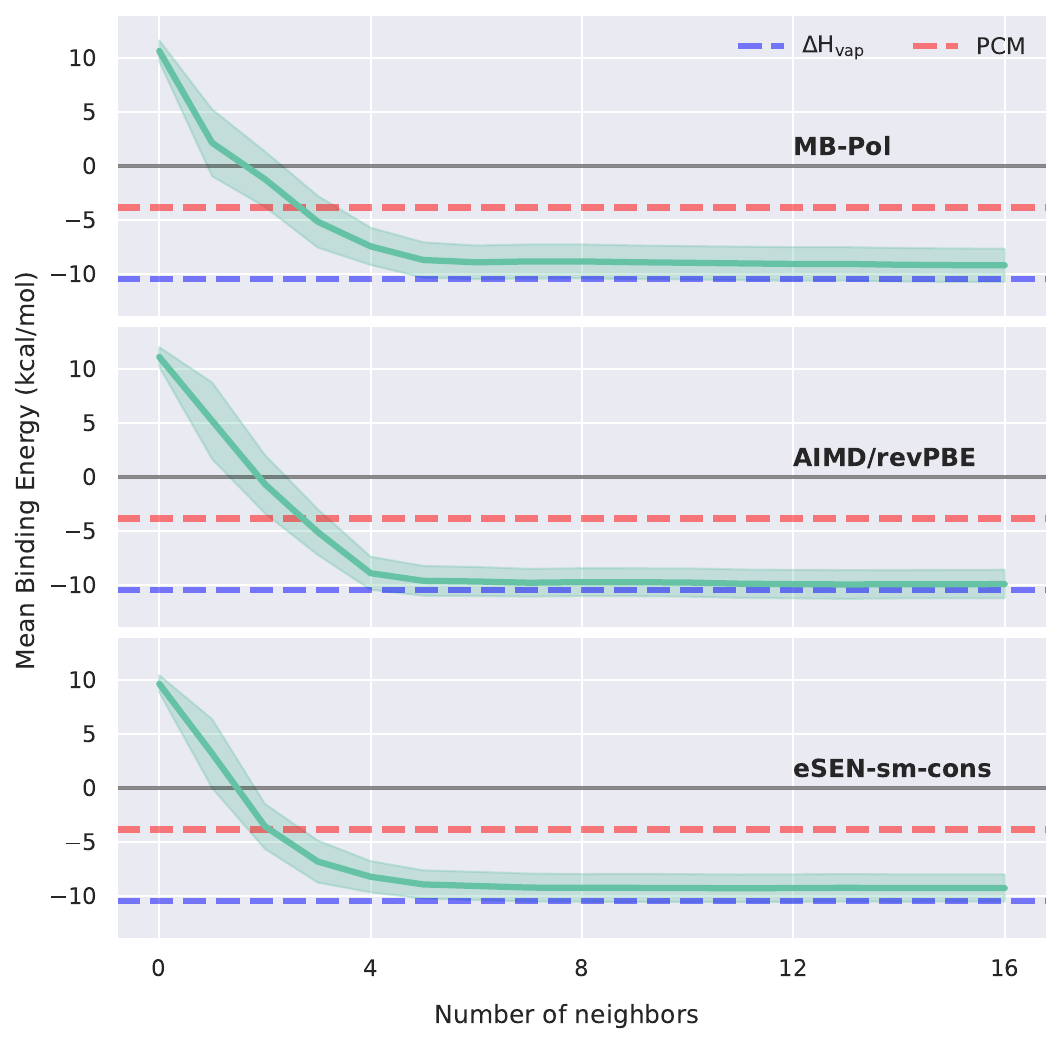}
    \caption{Comparison of different water samplings for computing mean binding energies on the basis of the decomposed approach in Eq. \ref{solv_energy_eq}. In all three sets of results, the level of theory is $\omega$B97M-V/aug-pcseg-1. For details on the individual MD simulations, please see Sect. \ref{comp_detail_sect}.}\label{comparison_sampling_fig}
\end{figure} 
Finally, we compare the influence of the underlying sampling behind Eq. \ref{solv_energy_eq} on our results for water. The results in Figs. \ref{comparison_eqs_fig_water} and \ref{comparison_panels_fig} have all been based on the MB-Pol MD sampling discussed in Sect. \ref{comp_detail_sect}. Under the assumption that this sampling can be used as a fair reference, the results in Fig. \ref{comparison_sampling_fig} (alongside the radial distribution functions in Fig. S2 of the SI) show that our proposed protocol for simulating binding energies is reasonably agnostic with respect to the sampling. Importantly, the computationally most efficient alternative---the {\texttt{eSEN-sm-cons}} potential---yields results using the present protocol that are practically indistinguishable from those of the MB-Pol sampling, whilst also being applicable to liquids other than water. For this reason, alongside the remarkable and transferable performance reported in Ref. \citenum{Fu2025}, we will base the results to follow for ethanol and acetonitrile on {\texttt{eSEN-sm-cons}} samplings.

\subsection{Ethanol and Acetonitrile}\label{ethanol_acetonitrile_res_subsect}

To study the performance of our new protocol across other types of homogeneous condensed phases, we will now apply it also to liquid ethanol and acetonitrile. The former of these two represents yet another popular solvent capable of forming hydrogen bonds, although not to the same extent as water, while the latter is an example of an aprotic solvent.
At a molecular level, alcohols are thus composed of both hydrophilic and hydrophobic groups~\cite{aagren_ethanol_jpcb_2014,aagren_ethanol_acr_2022}, while the important attractive interactions in acetonitrile clusters manifest from a combination of relatively weaker intermolecular C--H$\cdots$N bonds and long-range dipole-dipole interactions~\cite{Jorgensen1988, Takamuku1998, Mata2004, Remya2014}.\\

%
\begin{figure}[ht!]
    \centering
    \includegraphics[width=\textwidth]{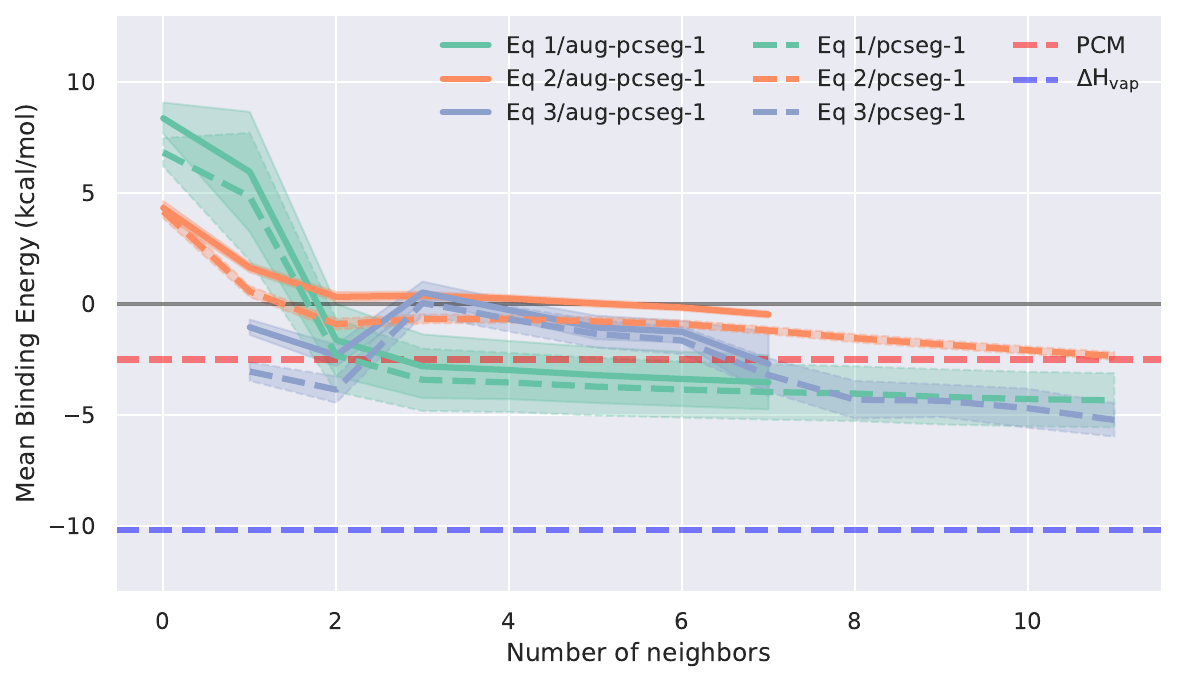}
    \caption{Results for ethanol on par with those reported for water in Fig \ref{comparison_eqs_fig_water}.} \label{ethanol_wb97mv}
\end{figure}
Similar to the results for the mean binding energy of water in Sect. \ref{water_res_subsect}, those for ethanol in Fig. \ref{ethanol_wb97mv} are observed to converge smoothly with respect to the number of surrounding neighbors when based on Eq. \ref{solv_energy_eq}, albeit not at quite the same rapid rate as for water. This is evident when comparing the actual convergence rates for the two systems in Figs. S5 and S6 of the SI. However, unlike for the earlier case of water, results are observed not to change much even in a minimal-like (pcseg-0) basis set, cf. Fig. S3, although the BSSE correction is significantly larger in this basis. The fact that ethanol only has a single hydroxyl group capable of forming hydrogen bonds to its neighbors is reflected also in the distributions of individual binding energies in Fig. S11 ($\mathcal{N}=200$, reported on par with Fig. \ref{swarmplot_mbpol_fig} for water).\\

The variation with choice of $xc$ functional is, however, observed to be significantly more pronounced. In Fig. S4 of the SI, results are once again compared using the B3LYP, MN15, and $\omega$B97M-V $xc$ functionals, all in a pcseg-1 basis set. Here, the B3LYP results even fail to predict a net stabilization of ethanol in the bulk phase, in contrast to both MN15 and $\omega$B97M-V. Earlier studies on computing binding energies for several (small and optimized) neutral ethanol clusters have found MN15 to agree reasonably well with higher-level coupled cluster~\cite{conradie_ethanol_pccp_2020}, with larger differences again observed in the case of B3LYP~\cite{yanez_ethanol_jcp_1999}. One key difference between the present work and those earlier studies, in addition to the fact that we here decompose energies rather than simply averaging out results (akin to Eq. \ref{average_energy_eq}), adheres to our sampling of the bulk dynamics, as opposed to geometry-optimized clusters, and, thus, implicitly how we account for temperature effects. As is evident from Fig. S11 of the SI, individual binding energies in liquid ethanol exhibit much the same degree of variation as was previously observed for liquid water in Fig. \ref{swarmplot_mbpol_fig}. In contrast to water, however, the individual contributions to $\mathcal{E}^{(n)}_{\mathcal{K}}$ of Eq. \ref{solv_energy_eq} are observed to converge slower, cf. Fig. S9 of the SI, and an approximate account of dispersion interactions, particularly that of the $\omega$B97M-V $xc$ functional, is found to be integral in order to obtain results in reasonable agreement with experiment, $\Delta \text{H}_{\text{vap}}$, although the difference is larger than what was observed for water. We speculate that this larger deviation is most likely due to the intrinsic limitations of KS-DFT.\\

\begin{figure}[ht!]
    \centering
    \includegraphics[width=\textwidth]{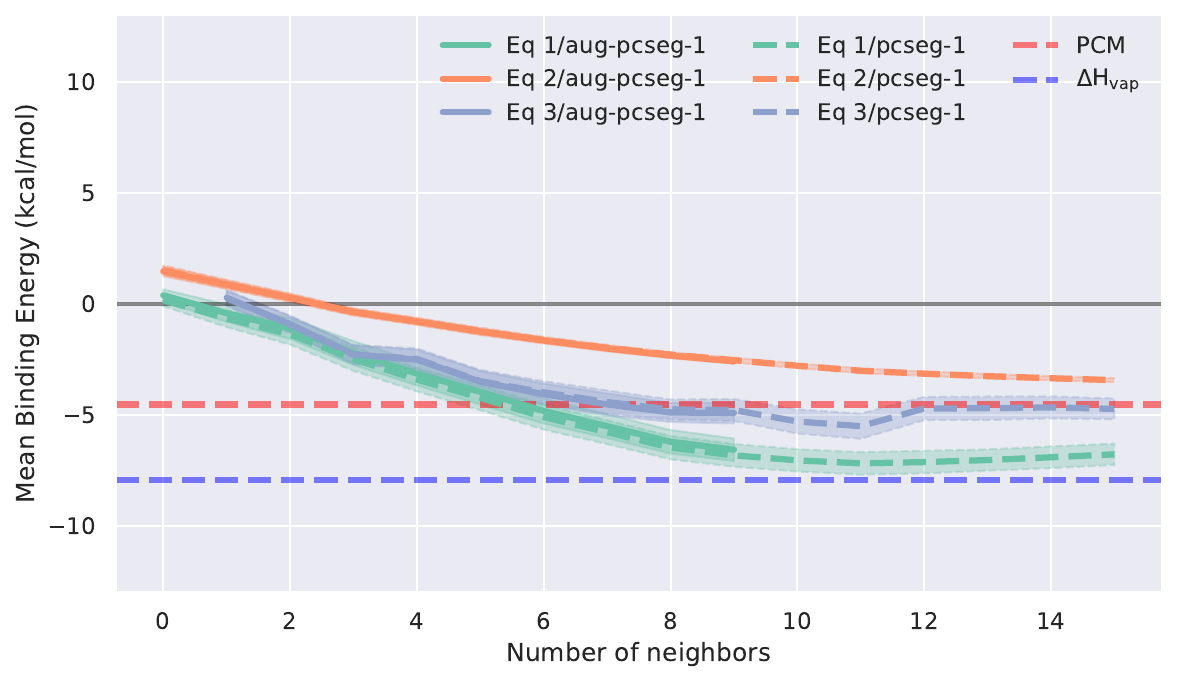}
    \caption{Results for acetonitrile on par with those reported for water in Fig \ref{comparison_eqs_fig_water}.} \label{acetonitrile_wb97mv}
\end{figure}
For acetonitrile in Fig. \ref{acetonitrile_wb97mv}, the lack of strong hydrogen bonds in the bulk is observed to give rise to significantly lower statistical uncertainties, which is, in turn, relates to the corresponding distributions in Fig. S12 of the SI being much narrower than for water and ethanol. The orientations of its neighbors do not affect the energy of a central acetonitrile monomer to the same extent, and, consequently, this system requires fewer uncorrelated configurations to reach the same error bars (although $\mathcal{N}=200$ snapshots were still used for consistency).\\

The convergence profile of the binding energy for acetonitrile is also observed to differ from those discussed for water and ethanol. For both of those systems, an initial destabilization precedes the eventual stabilization in the bulk upon adding a few neighboring monomers, whereas for acetonitrile, we instead observe a near-linear decrease in $\mathcal{E}^{(n)}_{\mathcal{K}}$ until convergence at about $n=8$. As is evident from Fig. S7 of the SI, both the vdW and BSSE contributions to the $\omega$B97M-V/aug-pcseg-1 binding energy in Fig. \ref{acetonitrile_wb97mv} are smaller in magnitude for acetonitrile than for the other two systems, while, once again, an approximate account of dispersion interactions, e.g., by means of the non-local VV10 treatment in $\omega$B97M-V, is crucial (cf. Figs. S4 and S10); as was the case for ethanol, the popular B3LYP $xc$ functional fails to predict a negative mean binding energy, while overall variations with basis set are once more observed to be minor (and less than what was observed in our initial application to water).

\section{Summary and Discussion}\label{sum_disc_sect}

In the course of the present work, we have introduced a new protocol for computing binding energies of individual monomers suspended in homogeneous condensed phases, with initial applications to liquid water, ethanol, and acetonitrile. The protocol, which decomposes solvated monomer energies from calculations on larger bulks, exhibits rapid convergence of the binding energy for all three systems with respect to the number of neighboring monomers in comparison to alternative approaches derived from the same set of constituent calculations.\\

In general, the binding energies computed herein show only very minor variations with respect to one-electron discretization, thus permitting the use of modest-sized rather than extended basis sets. However, and as expected, results are found to depend on the density functional approximation used to a much larger degree, thus reflecting how well these account for intricate bulk phenomena, such as, dispersion interactions. Overall, the $\omega$B97M-V $xc$ functional has been found to yield stable results for all three systems studied here that converge satisfactorily with the extent of the QM region used to decompose the binding energies.\\

We have observed excellent (in the case of water and acetonitrile) or, at the very least, qualitatively reasonable (for ethanol) agreement with the experimentally available property closest in nature to the binding energies of the present work, namely, the enthalpy of vaporization of each of these liquids. Unlike most other computational protocols, binding energies have been calculated directly from the bulk, that is, without recourse to explicit geometry optimizations, and we have thus accounted for temperature effects by means of sampling from MD trajectories. In a similar vein, our reference energies for the molecules in isolation have been thermalized prior to their transitions into the corresponding condensed phases.\\

The present protocol yields shifts to internal energies approximated at room temperature, i.e., estimates of energetic stabilization upon immersion. Experiments, however, typically probe changes to the free energies involved in the thermodynamic solvation process. Conceptually, alchemical free-energy perturbation (FEP) theory for a homogeneous liquid computes the self-solvation free energy as the excess chemical potential of a tagged molecule by first forming a non-transient cavity and then turning on interactions along a coupling parameter (cf. the classical split discussed in Sect. \ref{alternatives_subsect})~\cite{jorgensen_fep_wires_2014}. The instantaneous energy increment entering FEP---modulo the cavity work and reference-state conventions---may thus be viewed as the same object whose bulk average we compute by means of Eq. \ref{solv_energy_eq}. In principle, a cumulant expansion of a Zwanzig master equation for the coupled and interacting end-state would give our mean binding energy as the first cumulant, followed by a term proportional to the variance among the results in Figs. \ref{swarmplot_mbpol_fig}, S11, and S12 for the three systems~\cite{zwanzig_fep_jcp_1954}. However, these variance contributions are obviously substantial in our case, and, in addition, our protocol omits the work associated with the formation of a cavity, which must be obtained by growing only the repulsive (soft-core) part of the solute–solvent potential before switching on electrostatics and dispersion. For neutral self-solvation in homogeneous bulks, this leg is typically modest and well-behaved, but it is, nonetheless, required, thus making a true alchemical calculation of solvation free energies within our theoretical setting somewhat complicated~\cite{kollman_fep_chem_rev_1993}.\\

Moving forward, we are particularly interested in extending the type of decomposed formalism proposed herein to the study of heterogeneous systems instead, e.g., the stabilization of organic chromophores in solution, thus looking at microsolvation phenomena more generally and in greater detail than what is perhaps offered by calculations of changes to total (bulk) energies alone. Furthermore, applications beyond ground-state properties, e.g., simulations of excitation energies in solutions and how these may be partitioned into contributions from a central solute alone and its surrounding environment, will be the focus of future studies.\\

It is worth noting that an interesting protocol for computing solvation free energies from subsystem treatments within continuum solvation schemes has recently been proposed~\cite{neugebauer_gibbs_solv_jctc_2022}. Such approaches generally make use of PCM to account for the work required in the process of bringing a solute from a separate phase into an unperturbed solvent on account of a hybrid, cluster-continuum scheme~\cite{riveros_hybrid_wires_2020}. Here, one will need to account for the (configurational) entropy associated with the presence of the explicit solvent molecules, which is a non-trivial task on its own~\cite{kamerlin_warshel_fep_solv_cpc_2009}. As alluded to in Sect. \ref{sampling_subsect}, two main strategies for sampling the explicit solvent molecules exist: in addition to MD samplings, one could resort instead to the general technique of microsolvation~\cite{reiher_microsolv_jcc_2020,reiher_microsolv_jctc_2025,grimme_microsolv_jctc_2022,grimme_crest_jcp_2024}, before subsequently weighting (by means of a Boltzmann distribution for the probability) the structures at a given temperature. In this case, it might prove beneficial also to draw inspiration from more statistical perspectives on equilibrium thermodynamics~\cite{stein_stat_microsolv_jpca_2023}. Of the two strategies, the data sampled using MD, which aligns well with the approach opted for by us in here, were observed in Ref. \citenum{neugebauer_gibbs_solv_jctc_2022} to yield the most stable and reliable results for selected chemical reactions in solution. In order to compute converged solvation free energies from such a model, however, inclusion of more than just the first solvation shell was required, and it was noted how one must, in general, be prepared to handle on the order of close to 100 explicit solvent molecules in the QM region. To that end, it will be interesting to study the extent to which a decomposed protocol similar to the one proposed in the course of the present work will compare to such cluster-continuum schemes and, potentially, offer an advantage in terms of a reduction to the computational resources.

\section*{Acknowledgments}

J.J.E. acknowledges financial support from VILLUM FONDEN (a part of the VELUX FOUNDATIONS) under project no. 37411, the Independent Research Fund Denmark under project no. 10.46540/2064-00007B, as well as the Novo Nordisk Foundation under project no. NNF25OC0102951. F.L. acknowledges funding from the Italian Ministry of Research (PRIN 2022) under project no. 2022WZ8LME\_002.

\section*{Supporting Information}

The supporting information (SI) provides further details on the AO- and MO-based decomposition schemes of the present work in Sect. S1 as well as a number of additional results in support of the findings reported herein, cf. Sect. S2. Fig. S1 presents bootstrapping errors for the case of water, while Fig. S2 reports RDFs for the three water samplings of Fig. \ref{comparison_sampling_fig}. Next, Figs. S3 and S4 compare different basis sets and $xc$ functionals for the three different systems of Sect. \ref{res_sect} (water, ethanol, and acetonitrile), while Figs. S5--S10 compare the convergence of the individual contributions to Eq. \ref{solv_energy_eq} in more detail. Finally, Figs. S11 and S12 report individual binding energies for ethanol and acetonitrile on par with Fig. \ref{swarmplot_mbpol_fig}.

\section*{Data Availability}

Data in support of the findings of this study are available within the article and its SI.

\newpage

\providecommand{\latin}[1]{#1}
\makeatletter
\providecommand{\doi}
  {\begingroup\let\do\@makeother\dospecials
  \catcode`\{=1 \catcode`\}=2 \doi@aux}
\providecommand{\doi@aux}[1]{\endgroup\texttt{#1}}
\makeatother
\providecommand*\mcitethebibliography{\thebibliography}
\csname @ifundefined\endcsname{endmcitethebibliography}
  {\let\endmcitethebibliography\endthebibliography}{}


\begin{mcitethebibliography}{115}
\bibitem[Not()]{Note-2}
In contrast to the temperature-dependent vibrational contributions, the
  zero-point energy correction is not recovered through the MD sampling in our
  protocol and has therefore been omitted.\relax
\mciteBstWouldAddEndPunctfalse
\mciteSetBstMidEndSepPunct{\mcitedefaultmidpunct}
{}{\mcitedefaultseppunct}\relax
\EndOfBibitem
\bibitem[Not()]{Note-3}
Neighboring monomers are ranked on the basis of the shortest distance between
  any one atom and any of those constituting the central monomer of
  interest.\relax
\mciteBstWouldAddEndPunctfalse
\mciteSetBstMidEndSepPunct{\mcitedefaultmidpunct}
{}{\mcitedefaultseppunct}\relax
\EndOfBibitem
\end{mcitethebibliography}
\end{document}